\documentclass[traditabstract]{aa}

\usepackage{amsmath}
\usepackage{mathtools}
\usepackage{graphicx}
\usepackage{natbib}
\usepackage{multirow}
\bibpunct{(}{)}{;}{a}{}{,}
\usepackage[varg]{txfonts}
\usepackage{mathrsfs}

\usepackage{xspace}

\newcommand{\ergs}{\ensuremath{ \mathrm{erg\,s}^{-1} }}

\newcommand{\gra}{\raisebox{1ex}{\scriptsize o}}

\def\la{\mathrel{\mathpalette\fun <}}

\def\fun#1#2{\lower3.6pt\vbox{\baselineskip0pt\lineskip.9pt
  \ialign{$\mathsurround=0pt#1\hfil##\hfil$\crcr#2\crcr\sim\crcr}}}
{\catcode`\|=\active
  \gdef\Braket#1{\left<\mathcode`\|"8000\let|\bravert {#1}\right>}}
\def\bravert{\egroup\,\vrule\,\bgroup}

\def\gtrsim{\ensuremath{\; \buildrel > \over \sim \;}}
\def\gsim{\lower.5ex\hbox{\gtsima}}

\newcommand{\inte}{\textsl{INTEGRAL}\xspace}
\newcommand{\integral}{\textsl{INTEGRAL}\xspace}
\newcommand{\xmm}{\textsl{XMM-Newton}\xspace}
\newcommand{\pn}{\textsl{PN}\xspace}
\newcommand{\mos}{\textsl{MOS}\xspace}

\newcommand{\rxte}{\textsl{RXTE}\xspace}
\newcommand{\xte}{\textsl{RXTE}\xspace}
\newcommand{\pca}{\textsl{PCA}\xspace}

\newcommand{\hexte}{\textsl{HEXTE}\xspace}

\newcommand{\ibis}{\textsl{IBIS}\xspace}
\newcommand{\isgri}{\textsl{ISGRI}\xspace}

\newcommand{\bat}{\textsl{BAT}\xspace}
\newcommand{\xrt}{\textsl{XRT}\xspace}
\newcommand{\swift}{\textsl{Swift}\xspace}

\newcommand{\jemx}{\textsl{JEM-X}\xspace}

\newcommand{\xspec}{\texttt{XSPEC}\xspace}
\newfont{\mc}{cmcsc10 scaled\magstep2}
\newfont{\cmc}{cmcsc10 scaled\magstep1}

\newcommand{\eflux}{{\rm erg\,cm^{-2}\,s^{-1}}}

\newcommand{\rx}{RX\,J0440.9$+$4431\xspace}

\newcommand{\bgi}{\begin{itemize}}
\newcommand{\edi}{\end{itemize}}

\newcommand{\be}{\begin{equation}}
\newcommand{\ee}{\end{equation}}

\newcommand{\bea}{\begin{eqnarray}}                  %
\newcommand{\eea}{\end{eqnarray}}                    %

\newcommand{\beaa}{\begin{eqnarray*}}                %
\newcommand{\eeaa}{\end{eqnarray*}}                  %

\newcommand{\bgd}{\begin{description}}
\newcommand{\edd}{\end{description}}

\newcommand{\bgf}{\begin{figure}}
\newcommand{\edf}{\end{figure}}

\newcommand{\bgc}{\begin{center}}
\newcommand{\edc}{\end{center}}

\newcommand{\bgt}{\begin{tabular}}
\newcommand{\edt}{\end{tabular}}

\newcommand{\bge}{\begin{enumerate}}
\newcommand{\ede}{\end{enumerate}}

\DeclareRobustCommand{\ion}[2]{%
 \relax\ifmmode
 \ifx\testbx\f@series
 {\mathbf{#1\,\mathsc{#2}}}\else
 {\mathrm{#1\,\mathsc{#2}}}\fi
 \else\textup{#1\,{\mdseries\textsc{#2}}}%
 \fi}

\def \compmag {\textsc{compmag}\xspace}
\def \comptt {\textsc{comptt}\xspace}
\def \compst {\textsc{compst}\xspace}
\def \compps {\textsc{compps}\xspace}
\def \comptb {\textsc{comptb}\xspace}
\def \BMC {\textsc{bmc}\xspace}

\def\ktbb{kT_\mathrm {BB}}

\begin{document}

\title{\object{RX\,J0440.9$+$4431}: a persistent Be/X-ray binary in outburst}

\author{C. Ferrigno
        \inst{1}
         \and 
         R. Farinelli \inst{2}
          \and 
        E. Bozzo\inst{1}
         \and
        K. Pottschmidt\inst{3,4}
	\and
        D. Klochkov\inst{5}
         \and
        P. Kretschmar\inst{6}
      }

\authorrunning{C. Ferrigno et al.}
\titlerunning{Outbursts of RX\,J0440.9$+$4431}
   \offprints{C. Ferrigno}

\institute{ISDC Data Center for Astrophysics,  Universit\'e de Gen\`eve, chemin d'\'Ecogia, 16, 1290 Versoix, Switzerland\\
	\email{Carlo.Ferrigno@unige.ch}
	 \and Dipartimento di Fisica Universit\`a di Ferrara via Saragat 1, I-44100, Ferrara, Italia
	\and CRESST \& University of Maryland Baltimore County, 1000 Hilltop Circle, Baltimore, MD 21250, USA
	\and NASA Goddard Space Flight Center, Astrophysics Science Division, Code 661, Greenbelt, MD 20771, USA
	\and IAAT, Abt.\ Astronomie, Universit\"at T\"ubingen, Sand 1, 72076 T\"ubingen, Germany    
	 \and ESAC, ISOC, Villa\~{n}ueva de la Ca\~{n}ada, Madrid, Spain
          }

\date{Received ---; accepted ---}

\abstract
	{The persistent Be/X-ray binary \rx flared in 2010 and 2011 and has been followed 
	by various X-ray facilities (\swift, \rxte, \xmm, and \integral). We studied the source timing and spectral properties
	as a function of its X-ray luminosity to investigate the transition from normal to flaring activity and the dynamical properties of the system.
	We have determined the orbital period from the long-term \swift/\bat light curve, but our determinations of the spin-period
	are not precise enough to constrain any orbital solution. The source spectrum can always be described by a bulk-motion 
	Comptonization model of black body seed photons 
	attenuated by a moderate photoelectric absorption. At the highest luminosity, we 
	measured a curvature of the spectrum, which we attribute to a significant contribution of the radiation pressure 
	in the accretion process. This allows us to estimate that the transition from a bulk-motion-dominated flow to a
	radiatively dominated one happens at a luminosity of $\sim2\times10^{36}\,\ergs$. 
	The luminosity dependency of the size of the black body emission region  is found to be 
	$r_\mathrm{BB} \propto L_X^{0.39\pm0.02}$. 
	This suggests that either matter accreting onto the neutron star hosted in \rx 
	penetrates through closed magnetic field lines at the border of the compact 
	object magnetosphere or that the structure of the neutron star magnetic field
	is more complicated than a simple dipole close to the surface.}
	

\maketitle

\section{Introduction}
\label{sec:intro} 

LS~V~+44~17/RX\,J0440.9+4431 was classified as one of the rare persistent neutron star Be/X-ray binaries 
based on the outcomes of the first dedicated \xte monitoring campaign performed by \citet{reig1999}. 
The source distance was later estimated at ($\sim$3.3$\pm$0.5)~kpc by \citet{reig2005}. 
RX\,J0440.9+4431 is also reported in the seven-year and nine-year 
\integral all-sky survey \citep{krivonos2010,krivonos2012}. From the former to the latter catalog, the average flux of the source  
increased from ($0.95\pm0.15$)\,mCrab to ($4.0\pm0.1$)\,mCrab\footnote{In the energy band 17--60\,keV these correspond to 1.35 and $5.7\times 10^{-11}\,\eflux$, respectively.} 
(17--60\,keV) as it underwent several episodes of enhanced X-ray activity in 2010 and 2011 (see later in this section).  
The flux estimate given in the seven-year catalog is consistent with what has been measured previously by \citet{reig1999} and can thus be considered as the 
persistent level of the source. 
A similar flux ($0.9\pm0.5$)\,mCrab\footnote{In the energy band 15--150~keV this corresponds to $2.0\times10^{-11}\,\eflux$.}  
was also reported in the 54-month Palermo \bat Survey in the 15--150\,keV band \citep{cusumano2010}.

\begin{figure}
  \begin{center}
  \resizebox{\hsize}{!}{
      \includegraphics[angle=0]{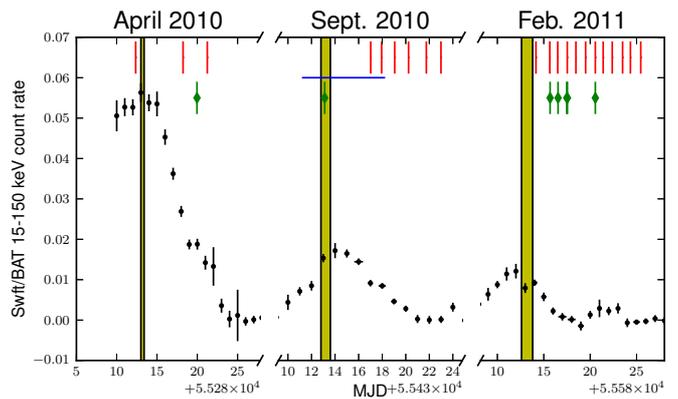}
	}
	\caption{Light curve of \rx and times of the observations reported in this paper. Black points with errors: \swift/\bat daily averaged light curve. Red vertical bars: times of the \xte observations. Blue horizontal bar: time span of the \inte data used in this paper. Green diamonds: times of the \swift observations used in our analysis. The first vertical yellow bar indicates the occurrence of the first maximum in the light curve; the others correspond to the times of the following maxima predicted by our orbital period determination. MJD 55\,280, 55\,430, and 55\,580 correspond to 2010, March 25, 2010, August 22, and 2011, January 19, respectively.}
\label{fig:lc}
\end{center}
\end{figure}

The first observed outburst from \rx, which we are aware of, was detected on 2010 March 26, and lasted until April 15 
\citep{morii2010, finger2010}. 
The \swift/\bat light  curve (Fig.~\ref{fig:lc})
showed that the source reached a peak luminosity of about 200\,mCrab (15-150\,keV) six days after the onset of the event. 
It remained in this state for eight days, and then turned off to its persistent luminosity level in about ten days.   
The outburst was also observed by the \xte All Sky Monitor (2-10 keV, \textsl{ASM}).
In the ASM light curve, the source reached about 10\,cts/s (i.e., $\sim$130\,mCrab).
A detailed study of the first outburst from \rx was presented by \citet{usui2011}. The authors show that the broad-band spectrum 
of the source could be described well by a combination of a black body component, 
a cut-off power law, and an iron line with a centroid energy of $\sim$6.4\,keV. 
They also find a pronounced dip in the pulse profile interpreted as being due to obscuration of the X-ray emission from the neutron star 
by the accretion stream.

A second outburst from \rx was detected by \integral on 2010 September 1 (Fig.~\ref{fig:lc}). 
The measured average and peak fluxes during the event were $\sim$30\,mCrab and $(77\pm17)$\,mCrab (17--60 keV), 
respectively \citep{krivonos2010b}. The \swift/\bat and \xte/\textsl{ASM} light curves 
showed that the outburst had a rise and decay time of about five days, with a compatible peak luminosity 
of $\sim$80\,mCrab. The analysis of the combined \swift, \inte, and \xte data led to identifying a possible 
cyclotron absorption feature at $\sim$30\,keV, suggesting a 
surface magnetic field of $B\simeq 3\times 10^{12}$\,G  for the accreting neutron star hosted in this system. 
This magnetic field intensity would be compatible with the measured 
properties of the source broad-band noise in the X-ray domain \citep{tsygankov2011}.

The two intense X-ray emission episodes recorded from \rx showed the typical properties of the so-called ``Type I'' outbursts, usually 
displayed when the neutron star in a Be/X-ray binary approaches the system periastron and interacts with the companion equatorial disk. 
The presence of a neutron star in \rx is secured by the detection of pulsations in its 
persistent and outburst emission. The measured pulse period increased 
from $\approx$202\,s in 1999 \citep{reig1999} to $\approx$206\,s during the latest observations, 
thus evidencing an average spin down trend. 
The first estimate of the source orbital period was reported by \citet{reig2011} at $\sim$150\,d
using the orbit-to-spin period relation for Be/X-ray binaries. 
This period would be compatible with the recurrence time of the source outbursts 
\citep[150--160~d][]{morii2010,krivonos2010b,tsygankov2011}. 

A deep (20\,ks) \xmm observation of \rx was carried out on 2011 March 18 to study the properties of its 
persistent emission \citep{lapalombara2011}. During this observation, the source displayed an average 
flux of 6$\times10^{-11}\,\eflux$ (0.3--12 keV) 
and the corresponding spectrum could be well described using a power law plus a black body model. 
The estimated radius for the thermal emission component 
was compatible with the size of a hot-spot on the surface of a strongly magnetized 
neutron star ($\sim$10$^{12}$\,G). Pulsations at $\approx$205\,s were detected 
throughout the observation. No significant spectral variability with the spin phase could be measured, although 
an indication of hardening was found at the pulse minimum.

In this paper, we report on the last detected outburst from the source which occurred on February 2011 (Fig.~\ref{fig:lc}). 
On this occasion, we triggered our
pre-approved \xte target of opportunity observations and performed a simultaneous monitoring campaign 
at the softer X-ray energies with \swift/\xrt to study the evolution of the source spectrum with luminosity and improve the 
measurement of the source orbital period. 
We have also re-analyzed all available \rxte, \swift/\xrt, \xmm, and \integral  \citep{rxte,xrt,xmm,integral} 
observations performed in the direction of \rx in 2010 and 2011 to compare the source emission properties 
across two orders of magnitude in luminosity. 
The data analysis technique is described in Sect.~\ref{sec:data}, and all the results are reported in 
Sect.~\ref{sec:results}. Our discussions and conclusions are given 
in Sect.~\ref{sec:discussion} and Sect.~\ref{sec:conclusions}, respectively.   

\section{Data analysis}
\label{sec:data}
We have analyzed \xte and \swift data using the {\sc heasoft} software package (v. 6.12) while all spectral fits have been performed using \xspec version 12.6.0u \citep{xspec}.

For \xte, we have followed the suggested analysis threads\footnote{\scriptsize\texttt{http://heasarc.gsfc.nasa.gov/docs/xte/xhp\_proc\_analysis.html}}
and we considered only \pca\ \citep[2--60\,keV, ][]{pca} data from the PCU2 unit, which was always active. 
Good Time Intervals (GTI) were created by imposing that 
at least 30 minutes passed after the satellite exited the SAA, and the elevation on the Earth limb was $\geq$10\gra.  
We additionally imposed an electron ratio $\leq$0.15 and modeled the background using the provided estimator 
v3.8. Events were extracted in the mode \texttt{GoodXenon} (time resolution 0.95\,$\mu$s 
and 255 energy channels). The instrument specific response matrices were generated taking into account the corrected 
energy bounds and the gain corrections performed by the \textsl{EDS}.  
The average \pca spectrum was obtained from the \texttt{standard2} mode without further rebinning and fitted above standard channel 4 ($\sim$3\,keV) 
up to the higher energy at which the source signal was detectable (i.e., about 20\,keV, for short observations or faint source states, higher otherwise). 
Spectra were assumed to have a systematic uncertainty of 0.5\%, as recommended by the instrument team\footnote{\scriptsize\texttt{http://heasarc.gsfc.nasa.gov/docs/xte/pca/doc/rmf/pcarmf-11.7/}}.

For \swift/\xrt (0.5--11\,keV), we followed standard analysis procedures \citep{burrows2005} and considered both Window Timing (WT) and Photon Counting (PC) 
data, but limited our spectral range to 0.8--10\,keV to avoid calibration uncertainties.  
\xrt spectra were grouped to have at least 50 counts in each energy channel using \texttt{grppha}.
We extracted separated spectra for data accumulated in each satellite orbit and used the proper exposure map to account for bad columns, hot pixels, 
vignetting, and PSF losses.

We extracted the source spectrum from the \integral \ibis/\isgri \citep[20--100\,keV,][]{ubertini03} and JEM-X 
\citep[5--30\,keV,][]{lund03} data of the 2010 outburst taken from 2010, September 1 to September 10 
using the Off-line Science Software (OSA, v.10) distributed by the ISDC \citep{isdc}.
The JEM-X spectrum is extracted in the standard 16 bins defined in OSA, while the IBIS/ISGRI one in user defined 62 bins.

\xmm data have been processed using XMM Science Analysis System (SAS) version 12.0.1.  
We have extracted only the \textsl{EPIC-PN} \citep[0.5-10\,keV,][]{pn}
spectrum from a region of 0.6 arcmin centered on the source. 
Background and pile-up corrections were neglected \citep{lapalombara2011}. 
The spectrum was adaptively re-binned using the SAS tool \texttt{specgroup} in order to have at least 25 counts per 
energy bin and, at the same time, to prevent oversampling of the energy resolution by more than a factor of three. 

\section{Results}
\label{sec:results}
\subsection{Orbital period}
\label{sec:bat}

\begin{figure}
  \begin{center}
  \resizebox{\hsize}{!}{
      \includegraphics[angle=270]{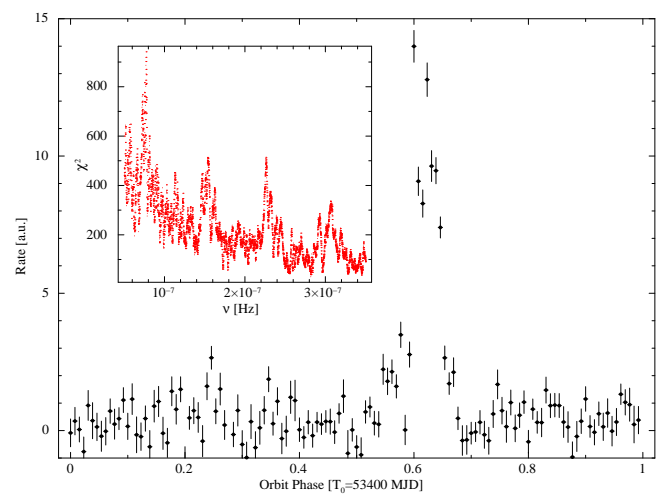}
	}
	\caption{Folded \swift/\bat light curve (15--150\,keV) of \rx at the best determined orbital period, $(150.0\pm0.2)$~days. 
	The inset shows the periodogram obtained from the epoch folding technique applied to the \swift/\bat light curve.}
\label{fig:epochfolding}
\end{center}
\end{figure}

To search for the source orbital period, we exploited the daily averaged light curve obtained from the hard X-ray transient monitor page of 
\swift/\bat\footnote{\scriptsize\texttt{http://heasarc.gsfc.nasa.gov/docs/swift/results/transients}}. These data cover almost continuously 
the time interval MJD~53\,414--56\,282 (2005 February 13 -- 2012 December 31), owing to the large field of view of the instrument and the spacecraft pointing strategy.
We first used the Lomb-Scargle periodogram method in the range 32--200\,days. 
The highest significance period is found at $\sim$150~days. 
We have also exploited the epoch folding technique on the same data by assuming 
32 phase bins and assigning weights of $1/\sigma^2$ to the individual bins (here $\sigma$ is the reported uncertainty on the 
source count rate). The most prominent peak ($\chi^2=942$) is found also in this case at $\sim$150~days
with evident harmonics at multiple frequencies 
(see inset in Fig.~\ref{fig:epochfolding}). 
We have refined our determination of the orbital period 
by limiting our search with the above methods around the most significant peak and using 130 phase bins. 
The best value is found at $(150.0\pm0.2)$~days.  
The uncertainty at 68\% c.l. on this measurement 
is obtained by bootstrapping 1000 light curves
with count rates randomly distributed within the uncertainty of the measurements, and 
then computing the standard deviation of the period determinations. 
We have folded the light curve at this period and obtained the profile shown in Fig.~\ref{fig:epochfolding}. 
This is reminiscent of what is usually observed from Be/X-ray binaries, with most of the emission concentrated 
during the periastron passage (phase 0.6 in the figure).  
We have fit the folded light curve with a constant plus a Gaussian to estimate the time of the light curve maximum. The Gaussian 
is centered at orbital phase $0.641\pm0.001$, which can be converted into the time of occurrence of the first flare by taking into account the 
best orbital period: it
corresponds to MJD~$55\,293.2\pm0.3$, i.e., the time
of the brightest flare observed so far from \rx. We have plot this time interval in Fig.~\ref{fig:lc}, left section, as a vertical shaded region. We have then propagated the
uncertainty on the orbital period to the following two flares observed in our campaign and plot them as shaded regions in the other two sections of the light curve. 

Interestingly, 
when limiting ourselves to the time range before MJD~55\,280 (2010 March 25), the orbital modulation was not detectable and no prominent peaks 
are clearly visible over the red noise. This indicates that the X-ray active phase of the source began only 
in April 2010 when it was detected in outburst for the first time, since the availability of modern X-ray monitors.

\subsection{Timing analysis and spin period measurements}
\label{sec:spin} 

To search for the $\sim200$\,s source spin period and its possible variations, we performed a timing analysis of all \rxte data (event mode \texttt{GoodXenon}).  
\swift pointings were on average too short to perform a similar analysis. 
We first extracted the background corrected \pca light curves in the 3--35\,keV energy range (time bin 2~s), 
and barycenter-corrected them using the known optical position \citep[]{reig2005}. For each observation we 
removed the variability on time scales longer than the expected pulsed signal\footnote{To achieve this, we computed 
the sliding average of the light curve on intervals of 400\,s every 200\,s and then
interpolated these values at the central time of each time bin to subtract them from the light curve. 
The resulting time series was then used as input to the epoch folding algorithm.} and determined the pulse period 
with the epoch folding technique. 

We obtained the best determined period for all the available \xte observations 
using an optimal number of phase bins (from 16 to 64 depending on the S/N), and estimated the 
uncertainties by running 100 simulations of the input light curves with rates estimated from Poisson 
distributions centered on the real measured values. The results are reported in 
Table~\ref{tab:periods}. Some variability in the measured values is evident from the table, but the limited number and accuracy of our
determinations are not sufficient  
to disentangle unambiguously the effects of the orbital motion and/or possible  
accretion torques.  

\begin{table}

\caption{Measurements of the spin period of \rx obtained from all the \xte observations analyzed in the present work.}
\begin{center}
\begin{tabular}{lccc}
\hline
\hline
Time &$\Delta T$\tablefootmark{a} & $P_\mathrm{spin}$ & $1\sigma$ error\\
 MJD & ks & s & s \\ 
\hline
55292.35 &  1.11 & 205.10 &   0.14\\
55298.23 &  1.24 & 205.03 &   0.13\\
55301.23 &  3.51 & 204.84 &   0.13\\
55447.02 &  3.60 & 204.89 &   0.17\\
55447.95 &  2.61 & 204.24 &   0.06\\
55449.06 &  3.32 & 205.51 &   0.07\\
55450.24 &  2.60 & 204.60 &   0.11\\
55451.75 &  3.67 & 205.38 &   1.19\\
55452.99 &  3.66 & 205.26 &   1.36\\
55594.16 &  3.67 & 204.65 &   0.29\\
55595.65 &  3.66 & 203.97 &   0.26\\
55596.50 &  2.79 & 205.42 &   0.19\\
55597.55 &  3.07 & 204.57 &   0.25\\
55598.45 &  2.95 & 205.00 &   0.18\\
55599.50 &  2.81 & 205.99 &   0.20\\
55600.55 &  3.35 & 205.78 &   0.25\\
55601.39 &  3.59 & 205.86 &   0.18\\
55602.37 &  3.56 & 206.29 &   0.18\\
55603.49 &  2.83 & 206.21 &   0.24\\
55604.34 &  0.97 & 205.25 &   0.86\\
55605.45 &  2.45 & 206.08 &   0.45\\
55606.49 &  1.48 & 206.01 &   0.47\\
\hline
\end{tabular}
\tablefoot{
\tablefoottext{a}{$\Delta T=\left(T_\mathrm{end}-T_\mathrm{start}\right)/2$.}
}
\end{center}
\label{tab:periods}
\end{table}

\subsection{Broad-band spectral analysis}
\label{sec:spectral}

The availability of quasi-simultaneous \integral, \xte, and \swift data permitted to investigate the broad-band
spectral properties of \rx \citep[note that part of the date were already presented:][]{tsygankov2011,usui2011}. 
The broad-band quasi-simultaneous data set is shown in Table~\ref{tab:summary} with a conventional \emph{ID} of the corresponding multi-instrument observations: A, S, 1, 2, 3, 4.
For the \integral data reduction we took advantage of the most updated \ibis/\isgri calibrations included in the 
OSA v.10\footnote{\scriptsize\texttt{http://www.isdc.unige.ch/newsletter.cgi?n=n23}}. 

To compare our findings with those reported previously by \citet{tsygankov2011}, and test the effect of the improved \integral calibrations, 
we used the same data set as in the above paper (\integral data from MJD~55440--55449, \xte/\pca 
data from OBSID~95418-01-03-00 and \swift/\xrt pointing OBSID~00031690003, i.e. observation S in Table~\ref{tab:summary}). 
We restricted the selection of pointings to a maximum off-axis angle of 7$^\circ$ 
to minimize the systematic uncertainties on the \ibis instrumental response. 
The total effective exposure time of the final \ibis/\isgri spectrum was 56\,ks. This corresponds to 
an effective exposure time for JEM-X2 of 64\,ks (owing to the lower dead-time of the detector). 

A spectral model comprising a cut-off power law 
and a black body component could fit the data 
reasonably well ($\chi^2$/d.o.f.=1.09/558). 
Normalization constants (fixed to 1 for \swift/\xrt) were also included in the fit to 
account for source variability and instrument inter-calibrations. 
Photoelectric absorption is accounted for
using the model \texttt{phabs} with abundances from \citet{anders1989}.
Some residuals are present at $\gtrsim$30\,keV, 
as reported by \citet{tsygankov2011}.  However, we found that an additional Lorentzian absorption 
feature with $\sigma$=6\,keV at $\sim$30\,keV, as proposed by these authors, did not significantly 
improve the fit ($\chi^2$/d.o.f.=1.05/556, 33\% chance improvement probability).

A better description of the spectrum could be obtained by using the 
using the general Comptonization model \BMC \citep{titarchuk1997}
with an exponential cutoff at high energy (\BMC$\times \exp{(-E/E_\mathrm{Fold})}$, $\chi^2$/d.o.f.=0.92/559). 
Because of the general shape of the convolutional kernel, 
this model can be applied to Comptonization spectra
of both optically thin and optically thick regimes, and considering
the cases in which thermal or thermal plus bulk Comptonization 
dominate \citep{farinelli2013}.
The emerging spectrum of the \BMC model can be expressed as 
\be
\label{eq:bmc}
F(E)=\frac{\mathrm{N_{BMC}}}{1+A}\left[ S(E) + A S(E) \ast G(E,E_0)\right]\,,
\ee
where $G(E,E_0)$ is the Green's function of the Componization energy operator, and $S(E)$ is the seed 
photon black body spectrum. 
The quantity $A/(A+1)$ empirically provides the fraction of seed photons which
are Compton scattered; it is related to the geometrical configuration
of the system and to the  seed photon spatial distribution in the
Comptonization region. Actually, $\log_{10}(A)$ is the model parameter returned
by \xspec. The Green's function $G(E,E_0)$ is a broken powerlaw
$\propto E^{\alpha+3}$ for $E < E_0$ and $\propto E^{-\alpha}$ for $E > E_0$.
The value of the powerlaw energy index $\alpha$  is related to
the efficiency of the Comptonization process.
From the normalization of the  model, it is possible to infer the effective radius of the region in which the 
seed black body radiation is generated by using the relation $\mathrm{N_{BMC}}$=$L_{39}/D_{10}^2$ (here  $L_{39}$ is the 
luminosity in units of $10^{39}$\,erg/s and $D_{10}$ the distance in units of 10\,kpc):  
\begin{equation}
r_\mathrm{BB,km}=\sqrt{7.75\times10^3\frac{ D_{10}^2 \mathrm{N_{BMC}}}{C{\left(kT\right)}_\mathrm{keV}^4}}\,. 
\label{eq:r_0}
\end{equation} 
Here $C$=1 for a spherical geometry and $C$=0.25 for a circular slab. Assuming for \rx a distance of 3.3\,kpc, 
the relation above can be written as $r_\mathrm{BB,km}=\sqrt{844\,\mathrm{N_{BMC}}/C{\left(kT\right)}_\mathrm{keV}^4}$.

\begin{figure}
  \begin{center}
  \resizebox{\hsize}{!}{\includegraphics[angle=270]{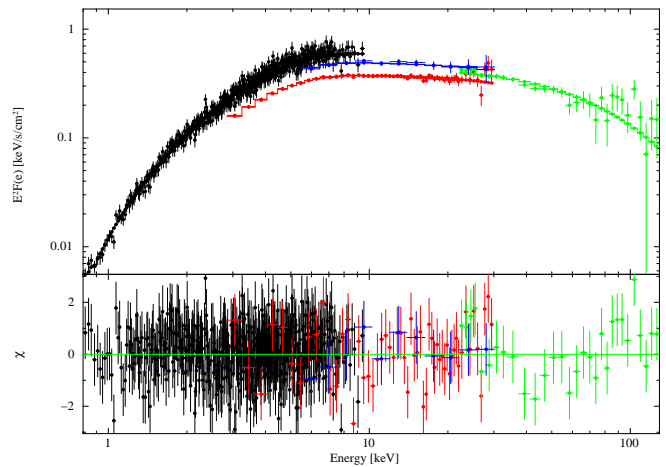}}
	\caption{Broad-band spectrum of the quasi-simultaneous \integral/\isgri (green), \integral/\textsl{JEM-X2} (blue), \xte/\pca (red), and \swift/\xrt (black) observations 
	performed in September 2010. The best fit is obtained with the absorbed \BMC with exponential cut-off model. The residuals from this fit are also shown 
	in the bottom panel.}
\label{fig:spec_integral}
\end{center}
\end{figure}

The best fit parameters obtained using the absorbed \BMC with exponential cut-off model are reported in Table~\ref{tab:fit_integral}.
We found that this description of the source continuum emission
permitted to constrain the optical depth 
of the previously suggested Lorentzian shaped absorption feature to $<$0.1 
(see the residuals from the best fit in in Fig.~\ref{fig:spec_integral}).  

To test further the applicability of the spectral model introduced above to \rx, we carried out a second fit to the quasi-simultaneous 
\swift/\xrt and \xte/\pca observations performed in April 2010 (observation A in Table~\ref{tab:summary}).
At this epoch, the source was brighter ($F_\mathrm{2-30\,keV}\simeq 3.8 \times 10^{-9}\,\mathrm{erg\,s^{-1}\,cm^{-2}}$) than in observation S, 
\pca data could be used from 3 to 45\,keV and an iron line was needed to  fit the data \citep[we modeled this as a thin additional Gaussian 
component centered at $\sim$6.4\,keV;][]{usui2011}.  
No evidence was found for the Lorentzian shaped absorption feature (width 6\,keV, centroid energy $<$42\,keV) suggested by 
\citet{tsygankov2011}. We obtained a 90\% c.l. upper limit of 0.03 on its optical depth.
We also note that \citet{usui2011} did not find evidence 
of any absorption feature in the combined \xte \pca and \hexte spectrum of the source. 
We conclude that the absorption feature reported by \citet{tsygankov2011} was probably due to systematic biases in the joint analysis 
of data from several instruments and/or a inadequate modeling of the continuum emission. 

\begin{table*}
\caption{Log of observations of the 2010-2011 outbursts used in this paper.}
\begin{center}
\begin{tabular}{cccccr@{}lcc}
\hline
\hline
\noalign{\smallskip}
\smallskip
ID\tablefootmark{a} & OBS ID\tablefootmark{b} & Exp.  & Start & Stop   &   \multicolumn{2}{c}{$\log_{10}($Flux$)$\tablefootmark{c}} & Flux\tablefootmark{c} & $L_X$\tablefootmark{d} \\
 &  & ks & MJD & MJD & & & $\times 10^{-9}$ & $\times 10^{35}$ \\
\hline
\smallskip
& 95418-01-01-00 & 3.1 & 55292.3306 & 55292.3778 & -8.2412 & $\pm 0.0007 $ & 5.738 & 74.766 \\
\multirow{2}{*} {A} & 95418-01-02-00 & 3.3 & 55298.2106 & 55298.2554 & -8.4228 & $\pm 0.0008 $ & 3.778 & 49.222 \\
\smallskip
 & 00031690001 & 4.8  & 55299.8793 & 55300.0424 & - & - & - & - \\
& 95418-01-02-01 & 3.2 & 55301.2178 & 55301.2611 & -8.7541 & $\pm 0.0013 $ & 1.762 & 22.954 \\
\hline
\smallskip
& 95418-01-03-01 & 1.0 & 55447.0197 & 55447.0311 & -8.738 & $\pm 0.002 $ & 1.827 & 23.804 \\
\multirow{4}{*} {S} & \ibis   & 56 & 55441.2576 & 55448.1785 & - & - & - & -  \\
& \jemx   & 64 & 55441.2576 & 55448.1785 & -8.78 & $\pm 0.01 $ & 1.66 & 21.6 \\
& 00031690003 & 5.7 & 55443.0294 & 55443.2021& - & - & - & - \\
\smallskip
& 95418-01-03-00 & 2.8 & 55447.9302 & 55447.9728 & -8.894 & $\pm 0.002 $ & 1.275 & 16.615 \\
& 95418-01-04-00 & 2.5 & 55449.0456 & 55449.0763 & -9.021 & $\pm 0.003 $ & 0.953 & 12.417 \\
& 95418-01-04-01 & 1.5 & 55450.2285 & 55450.2463 & -9.316 & $\pm 0.006 $ & 0.483 & 6.296 \\
& 95418-01-04-02 & 1.1 & 55451.7385 & 55451.7563 & -10.37 & $\pm 0.06 $ & 0.042 & 0.552 \\
& 95418-01-04-03 & 1.0 & 55452.9830 & 55452.9963 & -10.47 & $\pm 0.07 $ & 0.034 & 0.444 \\
\hline
\smallskip
& 96364-01-01-00 & 1.3 & 55594.1491 & 55594.1680 & -8.937 & $\pm 0.003 $ & 1.155 & 15.051 \\
\multirow{2}{*} 1 & 96364-01-01-01 & 1.5 & 55595.6400 & 55595.6574 & -9.120 & $\pm 0.004 $ & 0.759 & 9.888 \\
\smallskip
& 00031690005\tablefootmark{e} & 0.99 & 55595.6759 & 55595.6875 & - & - & - & -  \\
\multirow{2}{*} 2 & 96364-01-02-00 & 3.2 & 55596.4802 & 55596.5271 & -9.526 & $\pm 0.006 $ & 0.298 & 3.882 \\
\smallskip
& 00031690005\tablefootmark{e} & 1.46 & 55596.5314 & 55596.5484 & - & - & - & -  \\
\multirow{3}{*} 3 & 00031690006\tablefootmark{e} & 0.86 & 55597.4755 & 55597.4855 & - & - & - & -  \\
& 96364-01-03-00 & 3.0 & 55597.4921 & 55597.5728 & -10.33 & $\pm 0.03 $ & 0.046 & 0.604 \\
\smallskip
& 00031690006\tablefootmark{e} & 1.08 & 55597.5394 & 55597.5520 & - & - & - & -   \\
& 96364-01-04-00 & 3.2 & 55598.4385 & 55598.4848 & -10.38 & $\pm 0.02 $ & 0.042 & 0.547 \\
\smallskip
& 96364-01-05-00 & 2.9 & 55599.4872 & 55599.5304 & -10.41 & $\pm 0.04 $ & 0.039 & 0.504 \\
\multirow{2}{*} 4 & 96364-01-06-00 & 2.8 & 55600.5345 & 55600.5763 & -9.362 & $\pm 0.004 $ & 0.434 & 5.659 \\
\smallskip
& 00031690009 & 1.62 & 55600.5450 & 55600.5638& - & - & - & - \\
& 96364-01-07-00 & 3.3 & 55601.3754 & 55601.4217 & -10.40 & $\pm 0.05 $ & 0.040 & 0.515 \\
& 96364-01-08-00 & 3.3 & 55602.3547 & 55602.4000 & -10.40 & $\pm 0.04 $ & 0.040 & 0.517 \\
& 96364-01-09-00 & 3.0 & 55603.4719 & 55603.5132 & -10.54 & $\pm 0.03 $ & 0.029 & 0.376 \\
& 96364-01-10-00 & 3.2 & 55604.3171 & 55604.3582 & -10.63 & $\pm 0.06 $ & 0.023 & 0.304 \\
& 96364-01-11-00 & 3.1 & 55605.4297 & 55605.4710 & -10.63 & $\pm 0.03 $ & 0.023 & 0.302 \\
\hline
\end{tabular}
\tablefoot{
The three vertical blocks separated by a horizontal line refer from top to bottom to April 2010, September 2010, and February 2011, respectively.
\tablefoottext{a}{ID is the conventional name used in our broad-band spectral analysis exploiting quasi-simultaneous observations.}
\tablefoottext{b}{OBS IDs as 9xxxx-01-xx-xx refer to \xte, 0003169000x to \swift, \ibis and \jemx to the \inte monitoring.}
\tablefoottext{c}{2--30\,keV absorbed flux is in units of $\mathrm{erg\,cm^{-2}\,s^{-1}}$, not reported for instruments operating in different bands.}
\tablefoottext{d}{2--30\,keV luminosity in erg/s assuming a distance of 3.3\,kpc.}
\tablefoottext{e}{different snapshots with the same OBS ID.}
}
\end{center}
\label{tab:summary}
\end{table*}

\begin{table}
\caption{Best fit parameters for the absorbed \BMC with exponential cut-off model for the broad band spectral fits of the 2010 outburst (uncertainties are at 90\% c.l.).}
\begin{center}
\begin{tabular}{lr@{}lr@{}l}
\hline
\hline
 & \multicolumn{2}{c}{April 2010\tablefootmark{a}} &  \multicolumn{2}{c}{September 2010\tablefootmark{b} } \\
\hline
\smallskip
$N_\mathrm{H}$ $\left[10^{22}\mathrm{cm^{-2}}\right]$ & 0.31 & $\pm0.02$ & 0.27 & $^{+0.03}_{-0.02}$ \\
\smallskip
$E_\mathrm{Fold}$ [keV] & 29 &$^{+4}_{-3}$ & 53 & $^{+16}_{-10}$\\
\smallskip
$kT_\mathrm{BB}$ [keV] &1.38& $\pm0.02$ & 1.37 & $^{+0.03}_{-0.04}$ \\
\smallskip
$\alpha$ & 0.50 & $\pm 0.07$ & 0.78 & $\pm0.12$ \\
\smallskip
$\log_{10} A$ & $>2.15$ & & 0.8 & $^{+0.2}_{-0.1}$ \\
\smallskip
N$_\mathrm{BMC}\,\left[\frac{L_{39}}{D^2_{10}}\right]$\tablefootmark{c}& 0.030 & $^{+0.003}_{-0.001}$ & 0.019 & $\pm 0.002$ \\
\smallskip
$r_\mathrm{BB}$\tablefootmark{d} [km] & 2.64 & $\pm0.12$ & 2.13 & $\pm0.16$ \\
\smallskip
$E_\mathrm{Gauss}$ [keV] & 6.33 & $^{+0.04}_{-0.02}$ & & \\
\smallskip
$A_\mathrm{Gauss}$ & (6.6 & $\pm1.3)\times10^{-4}$ & &  \\
\smallskip
$C_\mathrm{\pca}$ & 1.55 & $\pm0.02$ & 0.632 & $\pm 0.008$ \\
\smallskip
$C_\mathrm{\isgri}$ & & &0.73 & $\pm0.03$ \\
\smallskip
$C_\mathrm{JEM-X2}$ & & &0.80 & $\pm0.02$ \\
\smallskip
Flux$_\mathrm{2-30\,keV}$\tablefootmark{e} & 3.78 &$\pm0.01$ & 1.272 & $\pm0.009$ \\
\smallskip
Flux$_\mathrm{0.1-200\,keV}$\tablefootmark{f} & 3.4 & & 2.9 &  \\
\smallskip
$\chi^2_\mathrm{red}$/d.o.f. & 1.24&/571 & 0.93&/559 \\
\hline
\end{tabular}
\tablefoot{
\tablefoottext{a}{observation A in Table~\ref{tab:summary}.}
\tablefoottext{b}{observation S in Table~\ref{tab:summary}.}
\tablefoottext{c}{$L_{39}$ is the X-ray luminosity in units of $10^{39}$\,erg/s, $D_{10}$ the source distance in units of 10\,kph}.
\tablefoottext{d}{quantity derived from N$_\mathrm{BMC}$ and $kT_\mathrm{BB}$.}
\tablefoottext{e}{flux of the \xte/\pca observation in units of  $10^{-9}\,\mathrm{erg\,cm^{-2}\,s^{-1}}$.}
\tablefoottext{f}{estimated de-absorbed flux in units of  $10^{-9}\,\mathrm{erg\,cm^{-2}\,s^{-1}}$, normalized to the \swift/\xrt observations.}
}
\end{center}
\label{tab:fit_integral}
\end{table}

\subsection{Spectral variability}
\label{sec:variability}
\begin{figure*}
  \begin{center}
  \resizebox{\hsize}{!}{\includegraphics[angle=0]{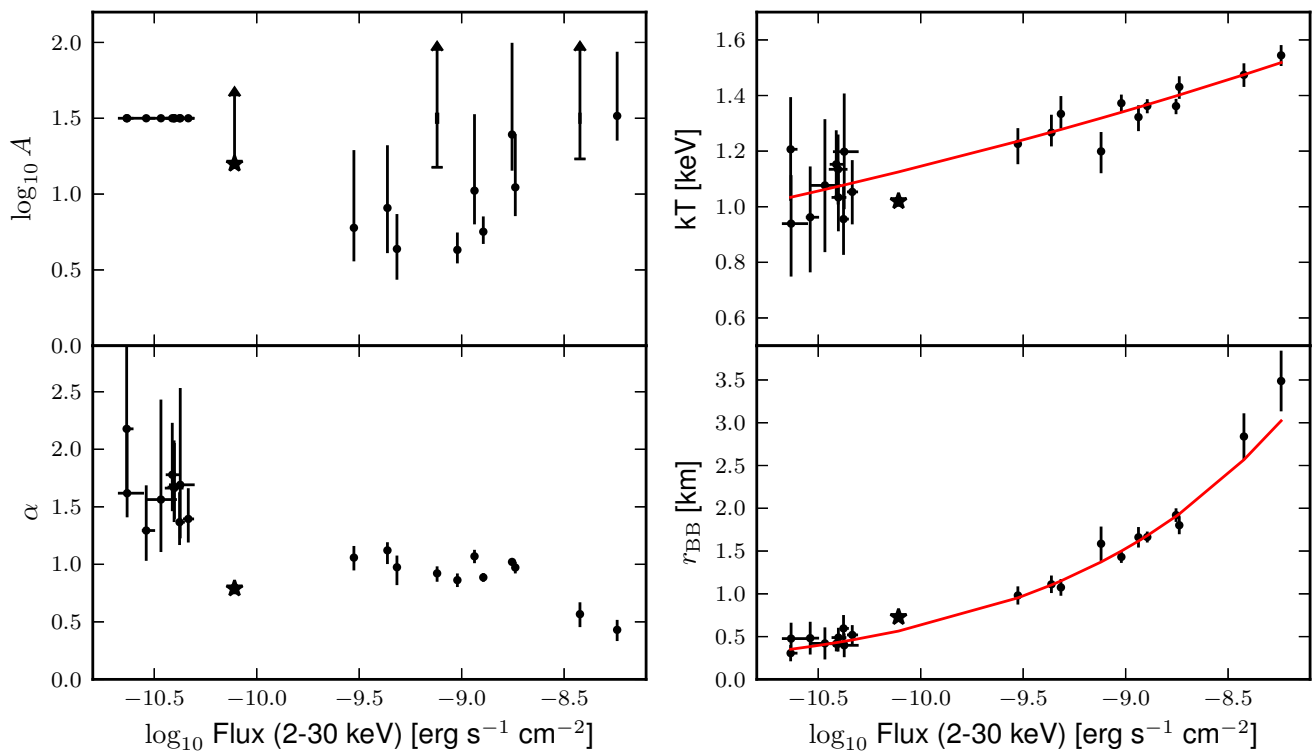}}
	\caption{Results of the spectral fits to the \xte/\pca observations (filled circles) performed during the outbursts 
	of \rx occurred in 2010 and 2011 
	(Table\,\ref{tab:summary}). The best fit spectral parameters obtained from the \xmm observation (2011 March 18) 
	are also included and represented with a star 
	(uncertainties are at 68\% c.l. and lower limits at 90\% c.l.).The \BMC model was used for all fits, but 
	during the two observations performed when the source reached the highest flux,
	an exponential cut-off at high energy 
	was needed to properly fit the data (see Sect.~\ref{sec:variability}).
	The black body radius $r_\mathrm{BB}$ is obtained from eq.~(\ref{eq:r_0}) setting $C=1$. 
	The red solid lines in the right panels represents the best fit relations of Eq.~(\ref{eq:r}--\ref{eq:kt}) obtained from the 
	\pca data only. }
\label{fig:refit}
\end{center}
\end{figure*}

To study possible changes in the spectral parameters of the source at different luminosity states, 
we fit here all the available \xte/\pca data, which provided an unbiased monitoring of the source across all different states,
by assuming the continuum model for the source broad-band emission 
discussed in Sect~\ref{sec:spectral}. A log of the observations is given in Table~\ref{tab:summary}, where we 
report also the source flux and luminosity.
The lower energy threshold of the \xte/\pca is 
3\,keV and it does not permit a precise measurement of the absorption column density in the direction of \rx.
Thus, we have fixed it
to the weighted average value measured from all the available \swift/\xrt and \xte quasi-simultaneous observations
characterised by high enough S/N to significantly constrain our parameter of interest 
(the observations with a conventional ID in Table~\ref{tab:summary}). 
We find 
$N_\mathrm{H}=(2.8\pm0.1)\times 10^{21}\,\mathrm{cm^{-2}}$. 
In several fits the value of $\log_{10} (A)$ in the \BMC model could only be poorly constrained.  
Where possible, we estimated a lower limit at 90\% c.l. by using the \xspec command \texttt{steppar}. 
For the lower statistic spectra we fixed it to $\log_{10} (A)=1.5$. 
During the observations performed when the source reached the highest flux 
(observation IDs  95418-01-01-00 and 95418-01-02-00 , April 2010) an exponential cut-off at high energy 
(i.e., $\exp{(-E/E_\mathrm{Fold})}$)
was needed to properly fit the data: we have found $E_\mathrm{Fold}=(30\pm6$)\,keV and ($33\pm8$)\,keV, respectively. 
For all the other observations, we have verified that the lower limit on the cut-off 
energy is outside the energy range covered by our data ($\sim$50\,keV), 
consistently with the value derived using \integral/\isgri data (see Sect.~\ref{sec:spectral}).
The results obtained from all the fits are reported in Fig.~\ref{fig:refit}. 
We include in the figure also the EPIC-pn data from the \xmm observation that caught \rx in its persistent low state \citep{lapalombara2011}. 
Note that the \BMC model also provided a fully acceptable fit to these data ($\chi^2$/d.o.f.=1.1/183) and yielded $\log_{10} A$$>$1.2, 
$kT=(1.02\pm0.03)$\,keV, $\alpha=0.81\pm0.08$, $r_\mathrm{BB}$=$(0.73\pm0.04)$\,km, and a column density $N_\mathrm{H}=(5.4\pm0.2)\times 
10^{21}\,\mathrm{cm}^{-2}$ (uncertainties at 90\% c.l.). The absorption column density measured with \xmm/\textsl{EPIC-PN} was higher than 
that measured with \swift/\xrt. We verified that fitting the \pca data with this higher value of $N_\mathrm{H}$ would not 
affect the results discussed below.

\begin{figure}
  \begin{center}
  \resizebox{\hsize}{!}{
      \includegraphics[angle=0]{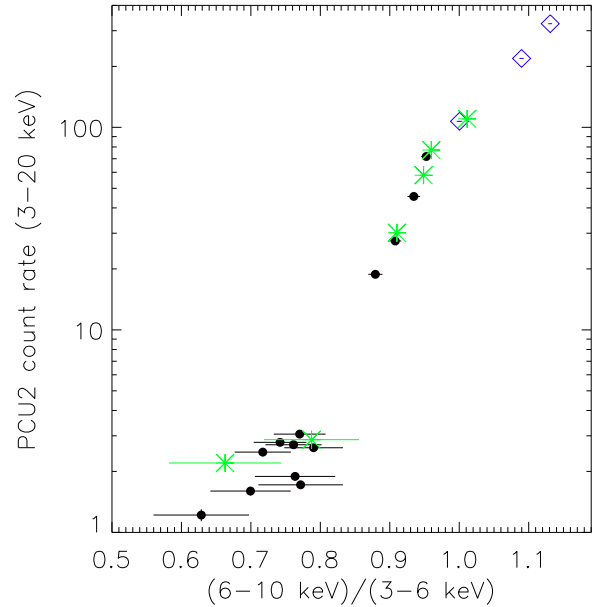}
	}
	\caption{Hardness-intensity diagram of all the available \pca observation of \rx. Different symbols refer to different outbursts: diamonds 
	are for the outburst in April 2010 (MJD\,55\,292--55\,301), stars for the outburst of September 2010 (MJD\,55\,447--55\,450), 
	and black dots 
	for the outburst of February 2011 (MJD\,55\,594--55\,606). 
	}
\label{fig:hid}
\end{center}
\end{figure}

From the plots in Fig.~\ref{fig:refit}, we note that the parameter $\log_{10} (A)$ could be only poorly constrained and $\alpha$ 
shows a not very pronounced downwards trend for increasing source flux.
Conversely, both the effective radius and temperature of the thermal emission component 
increase with the source flux. For the effective radius, we find ($\chi^2_\mathrm{red}$/d.o.f.=0.6/19):
\be
\label{eq:r}
\log_{10} r_\mathrm{BB} = (0.182\pm0.008) + (0.39\pm0.02)\left(\log_{10} F_\mathrm{-9}\right)\,,
\ee
where $r_\mathrm{BB}$ is the effective radius of the black body seed photons expressed in km and $ F_\mathrm{-9}$ is the 2--30\,keV flux 
expressed in units of $10^{-9}\eflux$. This relation is equivalent to $r_\mathrm{BB} \propto F_X^{0.39\pm0.02}$ or 
$F_X \propto r_\mathrm{BB}^{2.56\pm0.13}$. Similarly for the black body temperature, we find ($\chi^2_\mathrm{red}$/d.o.f.=0.6/19):
\be
\label{eq:kt}
\log_{10} kT_\mathrm{keV} = (0.128\pm0.004) + (0.070\pm0.008)\left(\log_{10} F_\mathrm{-9}\right)\,,
\ee
where $kT_\mathrm{keV}$ is the black body temperature expressed in keV. In eq.~(\ref{eq:r}-\ref{eq:kt}) uncertainties are given 
at $1\sigma$ c.l.
To prove  the variations in the source spectral properties in the different observations in a model-independent way,
we also report in Fig.~\ref{fig:hid} the hardness--intensity diagram of the \xte/\pca data \citep[as shown by, e.g.,][]{reig2013}. 
From this figure we note that during all outbursts the source X-ray emission becomes harder at higher fluxes and 
there is an evident change of the overall slope at PCU2 count rates $>$100 cts/s 
(the persistent emission level of \rx corresponds to a PCU2 count rate of about 1-3 cts/s)
We verified that different choices of the energy bands would not significantly affect this result.

An interpretation of the spectral changes in \rx is provided in Sect.~\ref{sec:discussion_variability}. 

\section{Discussion}
\label{sec:discussion}

\subsection{The long-term X-ray variability and the orbital period}
\label{sec:longterm}

The \swift /\bat data allowed us to measure with a fairly good accuracy the orbital period of the source owing to its enhanced X-ray activity 
in 2010 and 2011. Before this period, the available data revealed that \rx was a persistent X-ray emitter at a level of about 1\,mCrab in hard X-rays (15--150\,keV) 
and no significant orbital modulation could be detected from its light curve. Be/X-ray systems are rarely persistent X-ray sources, but many of them 
showed in the past only sporadic episodes of enhanced X-ray activity. The latter is thought to be originated by variations in the extension of the Be 
star equatorial disk \citep[see, e.g.,][]{reig2011}. The existence of persistent Be systems supports the idea that the fainter long-term X-ray 
activity of these sources is most likely related to the accretion of the NS from the tenuous wind of the companion, 
while the outbursts occur as a consequence of the 
interaction at the system periastron between the compact object and the Be equatorial disk \citep[see also,][, for the quiescent emission of A~0535+26]{rothschild2012}.
From the shaded regions of Fig.~\ref{fig:lc}, we note that the maxima of the second and third flares 
do not correspond exactly to a multiple of our best candidate orbital period: one lies slightly before, the other after. In Be/X-ray binary systems, 
some variability of the orbital phase occurrence of outburst is expected,  but in absence of 
orbital ephemeris based on pulse period modulation, we cannot speculate further on this subject.

\subsection{The broad-band spectral model}

Previous works on persistent Be pulsars \citep{reig1999} showed that their spectra can usually be well described by a model comprising 
a power law component and a black body with an emitting radius of 
a few hundreds meters \citep{lapalombara2006,lapalombara2007,lapalombara2009,lapalombara2011}. These sources are characterized 
by a spin period of the order of hundreds seconds, orbital period of hundreds of days and 
a persistent luminosity of $10^{34-35}\,\mathrm{erg\,s^{-1}}$. 
The direct black body emission is commonly interpreted as being due to the thermal mound at the base of the 
neutron star accretion column (with a size proportional to the width of the accretion stream), 
while the power law is the signature of Comptonization 
of the thermal photons by the in falling plasma. 
Following this interpretation and using data from several X-ray observatories, we have verified that also 
the broad-band spectrum of \rx could be reasonably well described by this phenomenological model, although our wide energy range
requirers the introduction of an exponential roll-over at high energies. 

A more insightful description of the parameters regulating the X-ray emission from the source can be obtained by using the \BMC model. 
This permits to describe in a self-consistent way the 
process in which part of the thermal seed photons are Comptonized and up-scattered to higher energies, giving rise to 
a power-law shaped emission with part of the photons escaping the Comptonizzation to reach directly the observer.  
The model is a general solution of the 
process of Comptonization of a spherically symmetric in-falling plasma but has the major limitation that the high energy roll-over,  
due to either the thermal motion of electrons or the deviation from a free-fall velocity profile of the moving plasma, is not taken into account. 
This can be empirically introduced by means of an additional exponential cut-off \citep[$\exp{(-E/E_\mathrm{Fold})}$,][]{farinelli2013}, still preserving 
the advantage of having a self-consistent treatment of the Compton up-scattering process. 

Several other models describe the process of Compton up-scattering of seed photons, 
with different degrees of complication and various limitations.
The models \comptt and \compst \citep[see][and references therein]{comptt} describe a thermal Comptonization
in a static plasma so, despite the possibility of obtaining statistically satisfactory results in terms of X-ray spectral
fitting, they are not suitable to be applied to accreting sources at highly sub-Eddington rates as \rx 
(in these cases the accreting material is expected to be nearly in free-fall).

The model \comptb \citep{farinelli2008} has been proposed 
to describe the emission of low-mass X-ray binaries for both the cases of simply thermal or thermal plus bulk Comptonization.
However, the correlation between the electron temperature $kT_e$, the spectral index $\alpha$ and the bulk 
parameter $\delta$ does not yield an independent determination of the whole parameter set. 
Assuming that bulk
Comptonization is an important effect, the model implementation requires that the parameter $kT_e$ is fixed or allowed to vary in a very narrow range, inferred from 
other considerations. It is also important to note that the \comptt, \compps, and \comptb models do not consider the presence of a magnetic
field so that the classical Thomson or Klein-Nishina electron scattering cross-sections are assumed.
The Be X-ray binary nature of \rx, together with its relatively high pulsed fraction \citep{usui2011}, 
suggests that the NS in this system is endowed with a magnetic field significantly higher than in the low-mass systems. 
This is expected to funnel efficiently the accreting material onto the neutron star's polar caps, leading to an electron temperature 
that is hardly predictable {\it a priori}.  

The model proposed by \citet{becker2007} is also not appropriate to describe the X-ray emission observed from \rx, as it 
assumes a velocity profile for the inflowing plasma in which an intense radiation field originates a collision-less shock 
above the NS surface. In this case, the radiation creeps through the lateral walls 
of the accretion stream giving rise to the so-called ``fan beam''. In the case of \rx, the emission pattern from the NS 
is expected to assume rather the shape of a ``pencil beam'', as at low luminosity  
the radiation is believed to be able to escape directly in the direction of the 
accretion stream and
no dramatic changes in the pulse profile could be revealed over two orders 
of magnitude in luminosity \citep{lapalombara2011,usui2011}. 
The latter assumption can be roughly demonstrated in the present case by comparing the magnitude of the gravitational force
of the neutron star and the radiation pressure from the polar cap where matter is thermalized. 
We find that
\begin{equation}
\mathscr{F}_{grav} \approx 1.3\times 10^{14} m {r_6}^{-1},
\end{equation}
\noindent
where $m$ and $r_6$ are the neutron mass in solar masses and the distance from the surface
in units of $10^6$\,cm, respectively.
On the other hand, the radiative force per unit mass is given by
\begin{equation}
\mathscr{F}_{rad} \approx 1.4\times 10^{13} {kT_\mathrm {bb}}^4 f(B),
\end{equation}
\noindent
where $\ktbb$ is the temperature of the thermal mound and
$f(B) \le 1$ is a correction factor to the Thomson cross-section due to the presence
of the magnetic field.
For photons propagating along the vertical direction, thus almost parallel to the magnetic field,
the Thomson cross-section is strongly reduced, with $f(B) \ll 1$ \citep{becker2007}.
For a typical neutron star mass and using the best fit values of 
$\ktbb$ reported in Tables~\ref{tab:fit_integral} and \ref{tab:fit_compmag}, it is
evident that $\mathscr{F}_{rad} \ll \mathscr{F}_{grav}$. 

\begin{table}
\caption{Best fit parameters of the \compmag\ model using a free-fall velocity profile of the accreting matter.}
\begin{center}
\begin{tabular}{lr@{}lr@{}l}
\hline
\hline
 & \multicolumn{2}{c}{April 2010\tablefootmark{a} } &  \multicolumn{2}{c}{September 2010\tablefootmark{b} } \\
\hline
\smallskip
$N_\mathrm{H}$ $\left[10^{22}\,\mathrm{cm^{-2}}\right]$ & 0.36 &$\pm$0.01 &  0.28 & $\pm$0.02 \\
\smallskip
$kT_\mathrm{BB}$ [keV] &1.50& $\pm0.01$ & 1.32 & $\pm$0.01\\
\smallskip
$kT_\mathrm{e}$ [keV] &13& $\pm 1$ &  62&   $\pm$1 \\
\smallskip
$\tau$ & 0.12 & $\pm$ 0.01 & 0.09 &  $\pm$ 0.01 \\
\smallskip
$\beta_0$ & & $>0.56$   &  0.44 & $\pm 0.01$ \\
\smallskip
$r_0$ & 0.30 & $^{+  0.05}_{-  0.02}$ & 0.32 & $^{+  0.01}_{-  0.04}$ \\
\smallskip
$N$ & 112 & $\pm 5$ & 388 & $\pm$1 \\
$E_\mathrm{Gauss}$ [keV] & 6.33 & $\pm 0.03$ & -- & \\
\smallskip
$A_\mathrm{Gauss}$ & (6.59 & $\pm1.2)\times10^{-4}$ & -- &  \\
\smallskip
Flux$_\mathrm{0.1-200\,keV}$\tablefootmark{c} & 3.4 &  & 3.0 &  \\
\smallskip
$\chi^2_\mathrm{red}$/d.o.f. & 1.21&/569 & 0.93&/558 \\
\hline
\end{tabular}
\tablefoot{
\tablefoottext{a}{observation A in Table~\ref{tab:summary}.}
\tablefoottext{b}{observation S in Table~\ref{tab:summary}.}
\tablefoottext{c}{estimate of de-absorbed flux in units of  $10^{-9}\,\mathrm{erg\,cm^{-2}\,s^{-1}}$, referred to \swift/\xrt observation.}
}
\end{center}
\label{tab:fit_compmag}
\end{table}

The model of \citet{becker2007} was extended recently \citep[\compmag;][]{farinelli2012a}. 
In this new implementation, it is possible to tune the velocity profile of the in-falling material. 
A broad-band spectral coverage is however necessary to obtain suitable constraints in the
physical parameters of \compmag, and thus the single \pca spectra could not be used for this purpose. 
More satisfactory results were obtained in the multi-instrument broad band spectral
analysis of the outbursts in April 2010 and September 2010. 
We have chosen to use a free-fall like profile with fixed parameter $\eta=0.5$, we have also fixed the 
albedo at the NS surface to one, while the other parameters are determined by the spectral fit.
They are the flow velocity at the NS surface ($\beta_0$, expressed in unities of 
$c$), the accretion column radius ($r_0$, expressed in Schwarzschild radii), the seed black body temperature ($kT_\mathrm{BB}$),
the electron temperature in the accretion stream  ($kT_\mathrm{e}$), their opticsl depth ($\tau$), and the model normalization ($N$). 
A summary of these results is reported in Table \ref{tab:fit_compmag}. 
The temperature of the seed thermal photons is in fair agreement with the 
\BMC model. The increase of the electron temperature at decreasing luminosity 
can be identified as the origin of the increased cut-off energy, while
the normalization, $N$,  of \compmag\ is defined such that $R_{\mathrm km}=N^{1/2} D_{10}$, and thus gives an estimate of 
the apparent blackbody emitting area. 
It is worth noting that \compmag assumes a uniform blackbody photon field  distribution
across the accreting column, so the inferred value  $R_{\mathrm km}$ does not actually
represent  the hot spot dimension on the neutron star surface, but rather
the total area obtained by emission at different heights above
the neutron star surface (so that in general $R_{\mathrm km} > r_0$).
This is confirmed here: we obtain $R_{\mathrm km} \sim 6.5$ and $R_{\mathrm km} \sim 3.5$
for the first and second observation, respectively (at a source distance of 3.3\,kpc), while the column radius 
remains constant within the uncertainties at about 1\,km.

We note that the absorbing column as measured by \xmm/\pn is significantly higher than the 
\swift/\xrt determinations, while the spectral parameters of the \BMC slightly deviate from the trend derived from the \pca data set (Fig.~\ref{fig:refit}). 
This can be due to 
the different energy ranges over which spectra are measured which can lead to a difference in the determination of the spectral slope and thus of
the absorbing column.
Furthermore, we note that according to \citet{romano2005}, the \pn detector yields spectral parameter for the calibration source PSR\,B0540$-$69 which deviate
from EPIC/\mos and \swift/\xrt in the same direction as our data, i.e., higher $N_\mathrm{H}$, although by a smaller amount. Therefore, we cannot 
exclude that the apparent discrepancy we have found is influenced by the calibration systematics. We have verified that using a different absorption model 
(\texttt{tbabs}) with the abundances of \citet{wilms2000} does not remove the discrepancy and yields an equally acceptable fit with 
compatible parameter value, except the absorbing column, which is systematically higher than in the model used in this paper.  
We remark that 
the absolute value of the absorbing column is of marginal importance for the results discussed here, and that the relatively low-resolution spectra analyzed here do not
allow us to distinguish between the models.
The values we have derived are similar
with the one determined by \citet{tsygankov2011}, while \citet{usui2011} use just \rxte data (above 3\,keV) and get absorbing column roughly one order of magnitude in excess:
we have verified that excluding the \swift/\xrt data, we would have obtained similar values, which we conclude to be inaccurate as they are hampered by the lack of low-energy coverage. 

Among all these different spectral models, the \BMC is able to provide a self-consistent and robust
physical description of the accretion process in \rx over the different luminosity levels even exploiting a limited energy band. 
The \compmag model is able to provide complementary information on the accretion process, which are not present in the \BMC model, although
with more structured assumptions on the geometrical properties of the accretion stream, in particular 
a significant vertical extent of the emitting plasma above the NS surface.
However, it generally requires a large energy coverage and it is of limited use when data are not available to provide constraints on  
the source emission above $\sim$20\,keV.

\subsection{Spectral parameters}
\label{sec:discussion_variability}
The observations of \rx in 2010 and 2011 allowed us to study spectral changes in the X-ray emission from the source 
over two orders of magnitude in X-ray flux and mass accretion rates ($\sim 10^{14-16}\,\mathrm{g\,s^{-1}}$).
\begin{figure}
  \begin{center}
  \resizebox{\hsize}{!}{
      \includegraphics[angle=0]{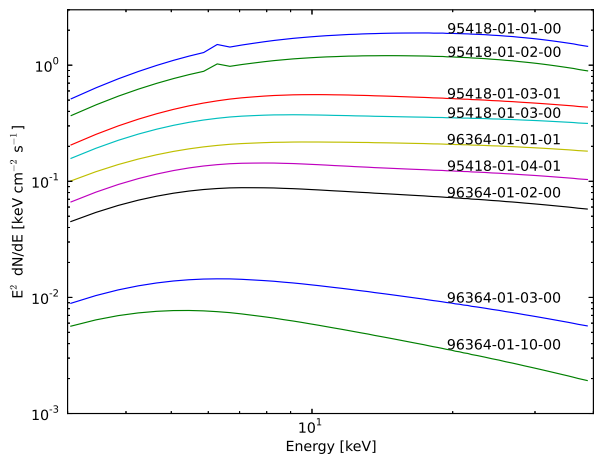}
	}
	\caption{A selection of unfolded \xte/\pca spectra at different luminosity states based on the \BMC model. 
	The corresponding observation ID is also indicated.}
\label{fig:eeuf}
\end{center}
\end{figure}
To compare the results of our spectral analysis at different accretion rates, we plot in Fig.~\ref{fig:eeuf} 
a selection of our \BMC best-fit unfolded spectral models obtained at increasing flux levels from the fits to the \xte data. 
We note that the spectra of the lower persistent state are rather soft, they become harder at intermediate states, and display a 
more pronounced curvature at the highest luminosities. 
This is in agreement with the measurement of a high-energy exponential cut-off at $\sim$30\,keV in the \xte/\pca data 
during the two observations performed when the source flux was $\geq10^{-9}\eflux$. The photon index 
of the Comptonization Green's function slightly decreases in comparison to the observations performed at the intermediate flux level, 
also indicating a hardening of the source X-ray emission. At the intermediate state, \integral/\isgri 
data were available to provide coverage at the higher energies, an exponential cut-off was measured at $\left(53^{+16}_{-10}\right)$\,keV. 
This suggests that the cut-off energy increases at lower luminosities so that the hard X-ray spectrum presents 
the hard tail expected when bulk Comptonization is the dominant reprocessing channel \citep{becker2005}.
Observations performed at the lower fluxes ($\leq10^{-10}\eflux$) 
showed a steeper Comptonization index and a cooler thermal photon temperature, they are thus characterized by a significantly softer emission. 
These results are confirmed by the model-independent analysis of the source emission provided by the 
HID in Fig.~\ref{fig:hid}. The observations in which a cut-off energy of 30\,keV was measured are represented with  
the two blue diamonds located in the upper right corner of the plot. The \pca observations performed at intermediate levels, for which 
quasi-simultaneous \integral/\isgri data were also available, are represented with the two green stars located immediately below the blue diamonds. 
The source emission hardens remarkabley at count rates $\sim$80--100\,cts/s, i.e., where the change in the cut-off energy is also revealed 
by the spectral fitting. The flux corresponding to the lowering of the cut-off energy   
is $1.3\div1.8\times 10^{-9}\,\eflux$, and  translates into a luminosity $\sim2\times10^{36}\,\ergs$. 
 
According to the discussion of \citet{becker2012}, at  low luminosity ($L_X\la10^{34-35}\,\ergs$) matter strikes the NS surface after passing through a gas-mediated shock. At higher luminosity, the radiation pressure begins to affect the accretion process so that
a radiative shock forms, and plasma deceleration takes place through Coulomb interactions in an vertically extended atmosphere. 
At this stage, the efficiency of the thermal cooling of the high-energy electrons responsible for the Comptonized emission
increases significantly
and can produce the observed change in the cut-off energy. However, the luminosity level at which this transition happens
is not well predicted on theoretical basis and is here tentatively identified on observational grounds at  $\sim2\times10^{36}\,\ergs$.
It is worth noticing that in all observations, \rx remains relatively faint in the X-rays compared to
the giant outbursts of transient Be/X-ray binaries ($L_X \sim 10^{38}\,\mathrm{erg\,s^{-1}}$). As a consequence,  
the HID does not show the typical upper branch turning towards softer colors typically observed  during the brightest outbursts 
of these sources, which might be interpreted as the onset of the fully radiation dominated accretion column \citep{reig2013,becker2012}.

\begin{figure}
  \begin{center}
  \resizebox{\hsize}{!}{
      \includegraphics[angle=0]{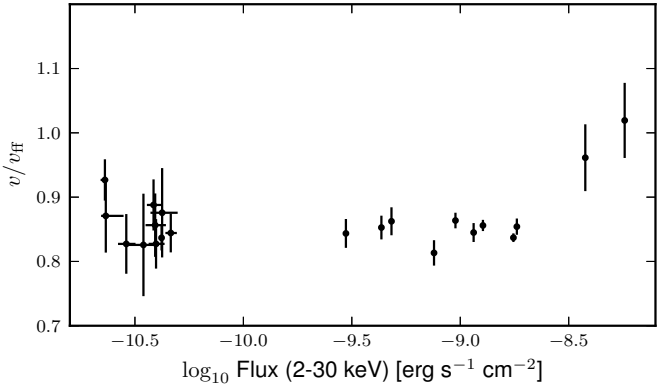}
	}
	\caption{Luminosity dependency of the velocity of the accreted matter at the NS surface obtained from eq.~\ref{eq:beta} assuming $C=0.25$.}
\label{fig:beta}
\end{center}
\end{figure}

To test further our understanding of the accretion processes in \rx given above, we have estimated from the \BMC best-fit parameters
the terminal velocity of the accretion 
flow at the NS surface for all the \xte spectra and reported our results in Fig.~\ref{fig:eeuf}.  
For this purpose, we follow eq.~(91--92) of \citet{becker2007} adapted to the notation used in the previous sections of our paper. 
By equating the energy flux dissipated by the thermally emitting surface to the flux of kinetic energy, we obtain 
\be
\sigma_\mathrm{SB} T_\mathrm{BB}^4 = \frac{1}{2}J v^2\,,
\ee
where $\sigma_\mathrm{SB}$ is the Stefan-Boltzmann constant, $T_\mathrm{BB}$ the black body temperature, $v$ the stream velocity, and
\be
J=\frac{\dot M}{\pi r_\mathrm{BB}^2}\,. 
\ee
The radius of the hot spot emitting the thermal radiation, $r_\mathrm{BB}$, can be derived from 
the measured spectral parameters using eq.~(\ref{eq:r_0}). We use also the standard conversion between 
mass accretion rate and X-ray luminosity:
\be
\dot M = \frac{4\pi D^2 F_X R_*^2}{G M_*}\,
\ee
where $D$ is the source distance, $F_X$ is approximated by $F_\mathrm{2-30\,keV}$, $G$ is the gravitation constant, 
$R_*$ and $M_*$ are the NS radius and mass. By taking into account that 
$\mathrm{N_{BMC}}=L_{39}/D_{10}^2$ (see eq.~(\ref{eq:r_0})) and expressing the free-fall velocity at the NS 
surface as $v_\mathrm{ff}=\sqrt{2G M_*/R_*}$, we obtain that 
\be
\frac{v}{v_\mathrm{ff}}=\sqrt{\frac{\mathrm{N_{BMC}}}{4.8\times10^7 C F_X}}\,,
\label{eq:beta}
\ee
where $C$ is the geometrical factor of equation~(\ref{eq:r_0}) and $F_X$ is expressed in $\eflux$.
By assuming that the thermal emitting surface is a circular slab ($C=0.25$), 
we derive a terminal velocity of the accretion stream which is $\approx0.85 v_\mathrm{ff}$ for all the observations, apart for the 
two corresponding to the highest luminosities (Fig.~\ref{fig:beta}).
In these cases the terminal velocity marginally exceeds the free-fall limit. This might indicate that the 
thermal mound is building-up on the NS surface so that the geometrical factor $C$ of the thermally emitting region 
increases and approaches that of the spherical geometry case ($C$=1). This indication agrees with the interpretation that the 
spectral change at the peak of the flaring activity of \rx is due 
to the formation of a radiatively dominated accretion column suggested by the decreasing cut-off energy. We note that the 
above calculations agree fairly well with the base velocity of the \compmag best-fit parameters ($\beta_0\approx 0.5 \approx 0.8 v_\mathrm{ff}/c$).

Throughout the whole luminosity range, the variation of the 
parameters of the thermal emision component in \rx could be satisfactorily described by the relations: 
$r_\mathrm{BB} \propto F_X^{0.39\pm0.02}$, $kT_\mathrm{BB}\propto F_X^{0.070\pm0.008}$ (see Sect.~\ref{sec:spectral} and Fig~\ref{fig:refit}). 
In the case of accretion onto a neutron star from the stellar wind, it is expected that the radius of the hot spot is 
inversely correlated to the luminosity, as at higher accretion rate matter should penetrate along magnetic field lines which 
are more and more concentrated towards the NS magnetic poles \citep{arons1980}. 
Conversely, if the transfer of matter is mediated by an accretion disk, the area of the hot spot
is proportional to the ratio $R_*/R_\mathrm{Alfven}$$\propto L_X^{2/7}$ \citep[see, e.g.,][]{lamb1973,white1983}. 
The correlation determined from the fits to the data thus support the idea that \rx is continuously accreting 
from a surrounding disk (despite the opposite would be expected for the persistent activity of a Be/X-ray binary, see Sect.~\ref{sec:longterm}).  
However, the measured relation between $r_\mathrm{BB}$ and the source luminosity is much 
steeper than expected in the case of disk accretion\footnote{To infer the thermal component radius from the 
model parameters, we have assumed a constant geometry (parameter $C$ of eq.~(\ref{eq:r_0})).}.
This might indicate either that matter penetrates through closed magnetic field lines at the border of the magnetosphere 
or that the neutron star magnetic field structure is a more complicated than a simple dipole close to the surface. 
The former idea was suggested to justify the magnetospheric instabilities observed 
in the outbursts of EXO~2030+375 \citep{klochkov2011} and A~0535+262 \citep{postnov2008}.
The latter has been invoked, e.g., by \citet{ferrigno2009} to interpret the soft
component in the broad-band spectrum of 4U\,0115+63 and can be related also to accretion-induced 
deformations of the magnetic field and instabilities in the accretion mound which are shown to exist on the base 
of recent numerical simulations \citep{mukherjee2012}.

The way in which the accreting matter penetrates the NS magnetosphere and accretes onto the compact star surface is also producing 
significant changes in its spin period. In Sect.~\ref{sec:spin}, we have shown that the spin period values measured from the observations of \rx 
are not precise enough to constrain the source orbital properties and/or  evaluate accretion torques.
However, we note that \rx showed a remarkable spin-down over the past 12 years from 202\,s \citep{reig1999} to $\approx$206\,s. 
If we assume that \rx is accreting from a disk at a non negligible rate, then 
the measured spin down can be interpreted in terms of coupling between the neutron star magnetic field lines and the external regions of the accretion disk 
located beyond the so-called corotational radius \citep{wang1995,wang1996}. This assumption has been sometimes used to constrain the NS magnetic field. 
As pointed out by \citet{bozzo09}, the uncertainties on the location and the extension
of the boundary layer in the disk accretion models for magnetized neutron stars are relatively large. 
Therefore, we could just derive that the magnetic field for the NS in \rx is of the order of few$\times$$10^{12}$\,G.  
This is in agreement with the timing signatures of the broad-band noise discussed in \citet{tsygankov2011}. 
The non-detection of the previously reported cyclotron scattering absorption line discussed in 
Sect.~\ref{sec:spectral} is not a counter argument to this magnetic field estimate,
as these features are not ubiquitous in the spectra of accreting magnetized neutron stars.

\section{Conclusions}
\label{sec:conclusions}

In this work, we reported on the analysis of the last outbursts detected from the persistent Be X-ray binary 
\rx, and used all archival observations of this source performed in 2010 and 2011 to investigate its spectral properties in 
a broad X-ray energy domain (0.3--150\,keV). 

We obtained the most accurate measurement of the source orbital period so far at $(150.0\pm0.2)$\,days and performed a detailed 
analysis of the source spectrum covering more than two orders of magnitude in luminosity. 
We used the bulk motion Comptonization model \citep[\BMC,][]{titarchuk1997}, which provided a self-consistent picture  
of the accretion processes in \rx. 
In particular, we found a coherent dependency of the size and temperature of the thermal emitting region as a 
function of the luminosity. These quantities do not follow the relations expected for accretion onto 
strongly magnetized neutron stars, and we argue that either accreting matter penetrates through closed magnetic field lines at the border of the magnetosphere 
or that the neutron star magnetic field configuration deviates from a dipole close to the surface. 

Analyzing the spectral changes at the hard X-ray band ($>$10~keV), we have identified the luminosity level at which the accretion flow 
dynamics begin to be affected by radiation pressure ($L_X\approx2\times10^{36}\,\ergs$). We estimated 
the in-falling plasma velocity close to the neutron star, which is consistent with predictions from a physically motivated model.
This provided an independent indication of the building-up of a radiative accretion column at the highest luminosity reached by \rx 
from the likely bending of the thermal emission surface.
\rx was particularly well suited for this study because the variations of its thermal emission could be unambiguously revealed 
during the higher and lower X-ray active states. This is not the case for other Be X-ray binaries \citep[see, e.g., the controversy on the spectral modeling of EXO~2020+375,][]{reig1999,klochkov2007,wilson2008}. 

We compared our findings on the accretion process in \rx by exploring the most recent spectral models accounting for 
thermal emission and its Comptonization in a strong magnetic field. 
The \compmag\ model could provide a number of further details 
for the interaction between the neutron star hosted in this source and the in-falling plasma, but the usability of this model is limited by the 
statistics of the available data. 
More sensitive missions will hopefully cover the required broad energy range (2--50\,keV) 
\citep[e.g., LOFT,][]{loft} and will allow us to study in greater detail the accretion process
at the lower end of the luminosity range of Be/X-ray binaries by means of complex physically motivated models,
which cannot be significantly constrained at present.

\section*{Acknowledgements}
This research has made use of data obtained through the High Energy Astrophysics Science Archive 
Research Center Online Service, provided by the NASA/Goddard Space Flight Center. We thank the \swift PI and the schedulers for 
having performed the target of opportunity observations used in this work in correspondence with the planned \xte TOOs. 
\bibliographystyle{aa}
\bibliography{lags,rx_lsv,exo_prop}

\end{document}